\documentclass[twocolumn,showpacs,preprintnumbers,amsmath,amssymb]{revtex4}
\usepackage{amssymb}
\usepackage{amsmath}
\usepackage{graphicx}
\usepackage{dcolumn}
\usepackage{bm}
\usepackage{color}
\usepackage{exscale}
\usepackage{relsize}
\usepackage{extarrows}
\usepackage{xcolor}

\begin{document}

\title{Quantum phase transition in the one-dimensional Dicke-Hubbard model with coupled qubits }
\author{Shu He$^{1}$}
\author{Li-Wei Duan$^{2}$}\email{duanlw@gmail.com}
\author{Yan-Zhi Wang$^{3,}$}
\author{Chen Wang$^{2,}$}
\author{Qing-Hu Chen$^{4}$}\email{qhchen@zju.edu.cn}
\address{
$^{1}$ Department of Physics and Electronic Engineering, Sichuan Normal University, Chengdu 610066, China \\
$^{2}$ Department of Physics, Zhejiang Normal University, Jinhua, 310018, China \\
$^{3}$ School of Physics and  Electronic Information, Anhui Normal University, Wuhu, 241002, China \\
$^{4}$ Department of Physics, Zhejiang University, Hangzhou 310027, P. R. China
 }

 \date{\today }

\begin{abstract}
  We study the ground state phase diagram of a one-dimensional two qubits Dicke-Hubbard model with XY qubit-qubit interaction. We use a numerical method combing the cluster mean-field theory and the matrix product state(MPS) to obtain the exact wave function of the ground state. When counter-rotating wave terms(CRTs) in the qubit-cavity coupling are neglected, we observe a rich phase diagram including a quantum phase transition between the Mott-insulating phase and the superfluid phase. This phase transition can be either the first-order or the second-order type depending on whether the total angular momentum changes across the phase diagram. Moreover, we observe two quantum triple points, at which three different phases coexist, with both positive and negative XY interactions. By further considering the effect of CRTs, we find that the main feature in the previous phase diagram, including the existence of quantum triple points, is retained. We also show that CRTs extremely demolish the non-local correlations in the coherent phase.
\end{abstract}

\maketitle

\section{Introduction}

As one of the most fundamental processes in nature, light-matter interaction plays a cornerstone role in our understanding of the quantum physics. The continued progress on the realization of elementary quantum building blocks with  unprecedented  experimental controllability, such as cavity QED{\cite{yoshie2004vacuum,birnbaum2005photon,fushman2008controlled,hennessy2007quantum}} and circuit QED{\cite{wallraff2004strong,nataf2010no,niemczyk2010circuit,buluta2011natural,houck2012chip}, has significantly boosted the research interest in the strong correlation and  collective behavior of photonic lattice systems.

One of the simple models to describe photonic lattice is the Jaynes-Cummings-Hubbard(JCH) model \cite{greentree2006quantum,hartmann2008quantum,schmidt2013circuit}. It is composed of  an array of cavities(or resonators) where each of them is individually coupled to a two-level atom(or qubits) and connected through the photon tunneling process. The competition between the intra-site qubit-cavity coupling and inter-site photon hopping leads to a quantum phase transition between the Mott-like insulating phase  which exhibits behavior similar to the photon
blockade\cite{birnbaum2005photon} and the superfluid phase in which photons are delocalized over the lattice\cite{hartmann2006strongly,angelakis2007photon,tomadin2010many,schiro2012phase}.

However, due to the achievement of strong and ultra coupling regime in the experimental realization of circuit QED\cite{bourassa2009ultrastrong,kockum2019ultrastrong},
the traditional Jaynes-Cummings description of qubit-cavity coupling under the rotating wave approximation(RWA) which neglects CRTs in the qubit-cavity interaction\cite{casanova2010deep}, is no longer valid. Then the Rabi-Hubbard(RH) model, where CRTs are fully considered, is introduced to correctly describe the nature of the quantum phase transition. Such CRTs reduce the $U(1)$ symmetry in the JCH model to the $Z_2$ symmetry and consequently prohibit the emergence of Mott lobes.  Previous studies demonstrate that the RH model undergoes a coherent-incoherent phase transition that belongs
to the universality class of the Quantum Ising model, which resembles the superradiant phase transition of the Dicke model\cite{chen2008numerically,schiro2012phase,rotondo2015dicke}.

Recently, some generalizations of the JCH and RH lattice models also attract great interest. In Ref\cite{cui2019two,cui2020nonlinear}, the two-photon process in the qubit-photon interaction is further considered in the original JCH  model. Unlike the one-photon model, a photon-pair superfluid phase where photons pair up in bound states is observed. In Ref\cite{schiro2016exotic}, the environmental dissipation is taken into consideration in the RH model and various exotic attractors are observed in the steady state phase diagram.

Another  extension to the RH model studies the collective effect induced by multiple qubits  embedded in each cavity, which is described by the so-called Dicke-Hubbard(DH) model\cite{lei2008quantum}. Previous theoretical study on the DH model demonstrates that the boundary of the coherent-incoherent transition is quantitatively affected by the number of qubits in a single cavity\cite{wu2017quantum}. Since multiple qubits are injected in a single cavity, the qubit-qubit interaction may not be negligible. In fact,the intracavity interaction among qubits has  been studied in the generalized Dicke model where dipole-coupled qubits interact with a single cavity mode\cite{chen2010quantum}. A phase diagram with both first- and second-order coherent-incoherent  phase transitions  are observed. The combination of the photonic lattice and qubit-qubit interaction, therefore, will undoubtedly enrich the  ground state feature and  quantum criticality.

In this paper, we use a numerically exact method combining the cluster mean field theory with the matrix product state  to study the quantum phase transition and ground state properties of the one dimensional Dicke-hubbard model with two dipole-coupled qubits in each cavity. Our study focuses on the following two aspects. Since a new degree of freedom is introduced by considering additional interatomic interactions, a natural question arises: how does this interaction affect the ground state phase diagram. As shown in the main text, quantum triple points emerge due to the qubit-qubit interaction. We will first study the properties of quantum phase transition around triple points in detail.  As one of the advantages of our numerical method,  the exact ground state wave function can be obtained. This allows us to explore the local and nonlocal properties of the ground state.

This paper is organized as follows. In Section \uppercase\expandafter{\romannumeral2}, we briefly introduce the one-dimensional Dicke-Hubbard model with dipole coupled qubits. In Section \uppercase\expandafter{\romannumeral3}, we present our main results consisting of three subsections: In Subsection A,  we study the effect of the qubit-qubit interaction on the phase diagram under the RWA.  Various observables are calculated to indicate the emergence of quantum triple points with corresponding first-order and second-order quantum phase transitions. In Subsection B, we discuss the local and non-local properties of different phases. In Subsection C, we generalize the model by considering CRTs and examine whether previous quantum triple points still exist on the phase diagram. In addition, we study the influence of CRTs on the non-local properties of the ground state. Finally, we present a summary in Section \uppercase\expandafter{\romannumeral4}.

\section{Models}
The Hamiltonian of the Dicke-Hubbard model describing an array of $M$ coupled cavities is given by:
\begin{align}
 &H = -J\sum_{m=1}^{M-1}(a_m^\dag a_{m+1} + a_m a^\dag_{m+1})   + \sum_mH_{\text{D},m}^M   \label{H1}
\end{align}
The first term in $(\ref{H1})$ describes the photonic tunneling process between adjacent cavities,and $J$ represents the tunneling strength. $a^\dag_m$($a_m$) is the photon creation(annihilation) operator of the $m$th  cavity mode.  The second term $H_{\text{D},m}$ describes the Hamiltonian of $K$ identical qubits coupled to  the single mode of $m$th cavity,
\begin{align}
 &H_{\text{D},m} = \omega a^\dag_m a_m + \frac{\Delta}{2} \sum_{i=1}^K\sigma^{z}_{m,i}  + {g_1}\sum_{i=1}^K \big(\sigma^+_{m,i}a_m +  \sigma^-_{m,i} a_m^\dag \big)\notag
 \\& +{g_2}\sum_{i=1}^K\big(\sigma^-_{m,i}a_m +  \sigma^+_{m,i} a_m^\dag \big) + \frac{\lambda}{4}\sum_{i\neq j}^K\big(\sigma^x_{m,i}\sigma^x_{m,j} + \sigma^y_{m,i}\sigma^y_{m,j}\big) \label{H2}
\end{align}
where $\omega$ is the frequency of cavity mode, $\Delta$ the qubit transition frequency, $g_1$ and $g_2$ the qubit-cavity coupling strength of rotating-wave terms  and CRTs respectively. $\sigma^{x,y,z}_{m,i}$ are Pauli matrices of the $i$th qubit in the $m$th cavity and $\sigma^{+}_{m,i}$($\sigma^{-}_{m,i}$) is corresponding raising(lowering) operator.  Note, we further take the qubit-qubit  interaction into account and assume that it has a form of XY interaction where $\lambda$ represents the strength of such interaction. In this case, the Hilbert space of the (\ref{H1}) is separated into two  noninteracting subspaces associated with even and odd $\mathcal{N}$ relatively. Under the RWA($g_1 =0$), two operators are commutative with the Hamiltonian (\ref{H1}). The first one is the  average excitation number $\hat{\mathcal{N}}$ defined as:
\begin{align}
&\hat{\mathcal{N}} =  \frac{1}{M}\left(\sum_n^M a^\dag_m a_m + \sum_{m,i}^{M,K}\sigma_{m,i}^+\sigma_{m,i}^-\right)
\end{align}
which has eigenvalues of $\mathcal{N} =0,1,2,...$. Another one is the average angular momentum operator $\hat{\mathcal{J}}^2$:
\begin{align}
& \hat{\mathcal{J}}^2 =  \frac{1}{M} \sum_m^M\left( J_{x,m}^2 + J_{y,m}^2 + J_{z,m}^2)  \right)\\
\end{align}
with eigenvalues of   $\mathcal{J}(\mathcal{J}+1)$ where $\mathcal{J} = K/2, K/2-1..., 0(1/2)$ when there are $K$ qubits in each cavity. Here $J_\alpha = \frac{1}{2}\sum_m \sigma_{\alpha,m}$ for $\alpha=x,y,z$ are the collective angular momentum operators.  For non-vanishing $g_2$, CRTs are not commutative with $\hat{\mathcal{N}}$. Thus the $U(1)$ symmetry is reduced to the discrete $Z_2$ symmetry with the parity operator $\hat{\mathcal{P}} = \exp(i\pi \hat{\mathcal{N}})$.

In the rest of the paper,  we study the quantum phase transition of the Hamiltonian(\ref{H1}) by using a numerically exact method that combines the matrix product state with the cluster mean-field expansion. The matrix product state method has been widely used to study the ground state properties of strong correlated one-dimensional lattice systems\cite{schollwock2011density,paeckel2019time}. Detailed implementation and convergence performance of this method are discussed in the Appendix. We restrict the number of qubits in each cavity $K=2$ and focus on the resonant case $\Delta=\omega=1$.

\section{Results and Discussion}

\subsection{Quantum phase transtions at $g_2 =0$}

\begin{figure*}[htp]
\centering
\includegraphics[scale=0.73]{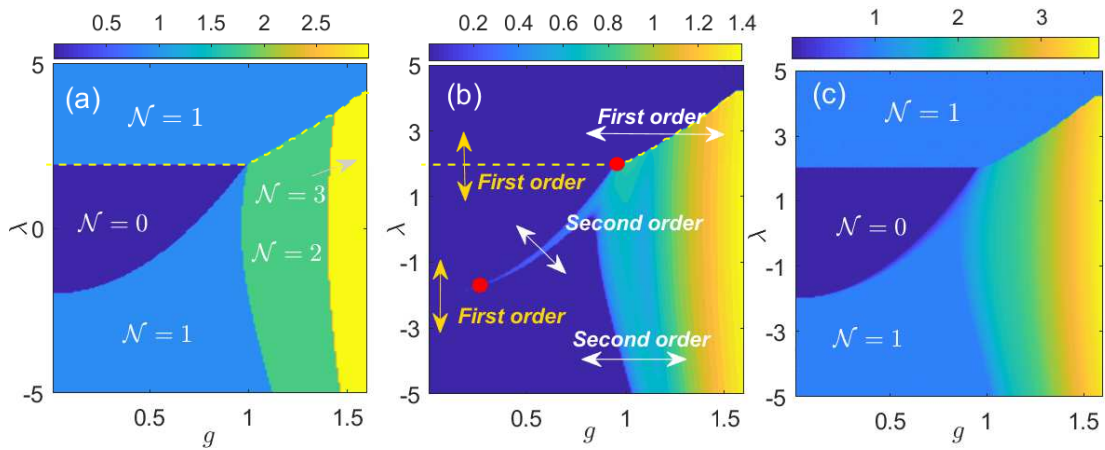}
\caption{(a) The phase diagram characterized by  the average excitation number $\mathcal{N}$  at $J=0$.  (b)The order parameter $\psi $ of ground state for various $g$ and $\lambda$ at  $J=0.05$.  Ground states with different angular momentum are separated by the yellow dashed line above(below) which $\mathcal{J}=0$($\mathcal{J} = 1$) (c) The average excitation number $\mathcal{N} $ as a function of $g$ and $\lambda$ at  $J=0.05$.}
\label{Fig1}
\end{figure*}

In this subsection, we study the ground state quantum phase transition of  the Hamiltonian (\ref{H1}) under RWA by setting $g_2 = 0$ and $g_1 =g$. Before discussing the main results, we first consider a special case that the hopping strength between two cavities is zero. As shown in Fig \ref{Fig1}.(a), we calculate the average excitation number $\mathcal{N}$ for various $\lambda$ and $g$. The ground states can be divided into several sectors  characterized by different integer $ \mathcal{N}$ due to the $U(1)$ symmetry and associated with a first-order quantum phase transition  between sectors. Since the angular momentum $\mathcal{J}$ is also a conserved quantity, we specify the ground state with different $\mathcal{J}$ by a yellow dashed line above(below) which $\mathcal{J} =0 $($\mathcal{J} =1 $).

Now we turn to the situation that $J$ is finite.  When there is no interaction between qubits ($\lambda=0)$, the Hamiltonian (\ref{H1}) reduces to the Dicke-Hubbard model with two qubits in each cavity. This model undergoes a second-order transition from the Mott-insulating phase to the superfluid phase  identified by the order parameter $\psi =|\langle a\rangle|$ where $a$ is the annihilation operator of the cavity mode. Considering  the cluster mean-field expansion used in our numerical method, we redefine the order parameter $\psi$ as the magnitude of the average value of the annihilation operator for each cavity mode in the cluster:
\begin{align}
  \psi = \left|\frac{1}{M}\sum_{m=1}^M\langle  \psi_{GS}|  a_m | \psi_{GS}\rangle\right|  \label{psiCMF}
\end{align}
In Fig \ref{Fig1}.(b) and (c), we display the order parameter $\psi$ and the average excitation number $\mathcal{N}$ for various $g$ and $\lambda$ at $J =0.05$.  Two quantum triple points, where three quantum phases coexist,  are observed at finite $\lambda$.  Associated quantum phase transitions can be classified into two categories: (1)Transition between the  Mott-insulating phase and the superfluid phase identified by the order parameter $ \psi = 0$ and $\psi\neq 0$  respectively.  (2)Transitions between  sectors with different integer $\mathcal{N}$ in the Mott-insulating phase.   To explore the nature of  triple points, we discuss the properties of these quantum phase transitions in detail below.

By comparing Fig \ref{Fig1}. (a) and (b), we find the boundary between the superfluid phase and the  Mott-insulating phase locates exactly along the critical line between sectors with different $\mathcal{N}$ at $J=0$ with two exceptions.  Firstly, the boundary between the region of $\mathcal{N} =1,\mathcal{J}=0$ and $\mathcal{N} =0,\mathcal{J}=1$ has no superfluid phase. The reason is explained as follows.  As we show in the next subsection, the correlation between two adjacent cavities is zero in the above two regions. Therefore, the ground state therein can be written as a product state of $\rho = \prod \rho_n = \prod |\psi_n\rangle\langle \psi_n |  $  where $|\psi_n\rangle $ is the ground state of the $n$th decoupled cavity described by the Hamiltonian (\ref{H2}). When $g=0$, it is straightforward to derive that  $|\psi_n\rangle =\frac{1}{\sqrt{2}}\left(|\uparrow \downarrow \rangle - | \downarrow \uparrow\rangle_{qb}\right)|0\rangle_{ph}  $ for $\lambda >2$ and  $|\psi_n\rangle =|\downarrow \downarrow \rangle |0\rangle_{ph}  $ for $ 2>\lambda >-2$.  Due to the constrain of the total excitation number,  such two simple forms hold for the whole region of $\mathcal{N}=1,\mathcal{J}=0$, $\mathcal{N}=0,\mathcal{J}=1$ respectively before entering into other sectors of higher $\mathcal{N}$ by increasing $g$. Both of their photon components are vacuum states which inactivate the photon hopping effect  and consequently prevent the emergence of the superfluid phase.  Secondly, there is no Mott-superfluid phase boundary between the region of $\mathcal{N}=2$ and $\mathcal{N}=3$ in Fig \ref{Fig1}. (a). The increase of $J$ will  enlarge the area of the Superfluid phase, especially at large coupling $g$ where more photons can be excited to exchange through hopping terms. As an example, we plot
the order parameter $\psi$ as a function of $g$ at different $J$. Fig \ref{Fig2}. (a). One can find that isolated areas of the superfluid phase characterized by nonzero $\psi$ gradually spread out around the phase boundary at $J = 0$(marked by black crosses) and finally merge  as $J$ increases to 0.06.

Note a small  part of the Superfluid phase imbeds into the Mott-insulating phase along the critical line that separates sectors of $\mathcal{N} =0$ and lower $\mathcal{N}=1$ , indicating the occurrence of two successive Mott-Superfluid-Mott phase transitions with the increase of  $g$. This overall pattern is quite similar to the phase diagram observed in the JCH model\cite{greentree2006quantum,schmidt2009strong,koch2009superfluid} where Superfluid phases are separated between isolated "Mott lobes" characterized by different polariton excitation numbers.

Unlike the JCH and the DH model where only second-order phase transitions are observed,   transitions between the Mott-insulating phase and the superfluid phase can be either the first-order or the second-order.   This can  be directly discerned by the first derivative of the ground state energy per site with respect to the coupling strength $g$. As depicted  in Fig \ref{Fig2}. (b).  $\partial E/\partial g$ displays an obvious discontinuity at the critical coupling $g_c$ for $\lambda = 2.2$ while a sharp but continuous cusp at $\lambda = 1.8$. A similar pattern is observed from other observables. This feature is in accordance with the criterion of the second- and first-order phase transitions. The key to understanding this difference is the angular momentum $\hat{\mathcal{J}}$.  Ground states in the region above the yellow dashed line in Fig \ref{Fig2}. (b) has conserved angular momentum with $\mathcal{J} =0$  , whereas $\mathcal{J} =1$ for the rest of the area. Therefore, with the increase of $g$,  $\mathcal{J}$ jumps from zero to one at the critical coupling $g_c$ when $\lambda >2$. It is this change of the angular momentum across the  phase boundary results in the unexpected first-order quantum phase transition between the Mott-insulating phase and the Superfluid phase.  Similar phenomenon has been previously observed in the Dicke model with coupled two-level atoms in a single cavity\cite{chen2010quantum}. Finally, by using the same criterion, we numerically confirm that the quantum phase transitions between different $\mathcal{N}$ in the Mott-insulating phase are all first-order.

\begin{figure}[htp]
\centering
\includegraphics[scale=0.63]{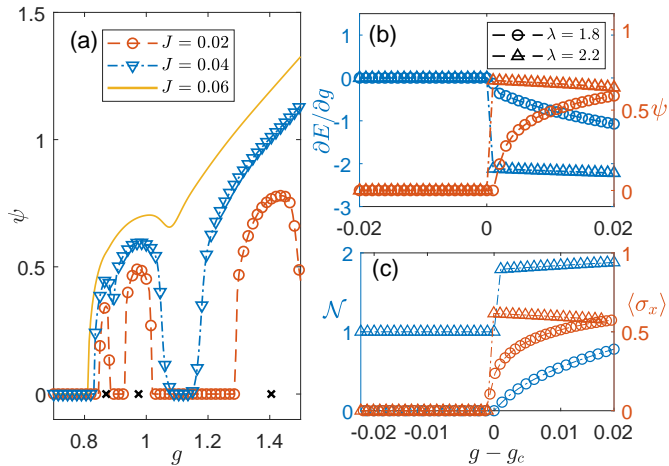}
\caption{
(a) The order parameter $\psi$ as a function of $g$ at different $J$. The black cross locates the phase boundary between different $\mathcal{N}$ at $J=0$.  (b),(c)The first derivative of the ground state energy with respect $g$ and  various observables including $\psi$, $\mathcal{N}$ and $\sigma_x$ near the phase boundary for $\lambda = 1.8$ and $\lambda = 2.2$.}
\label{Fig2}
\end{figure}

\subsection{ Properties of phases}
\begin{figure}[htp]
\centering
\includegraphics[scale=0.66]{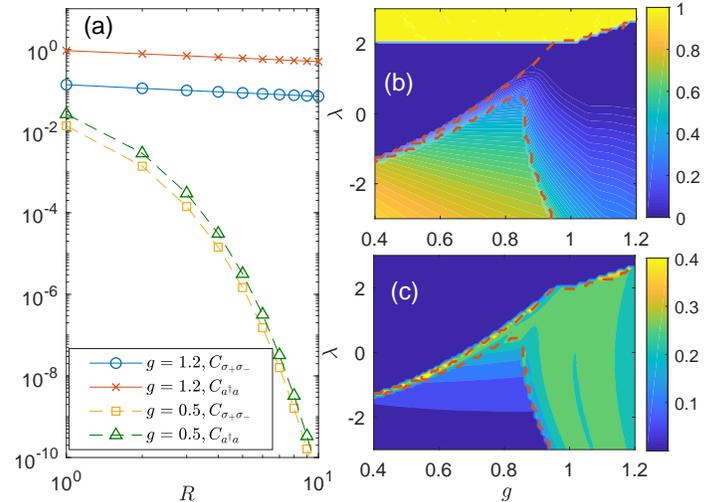}
\caption{
(a) The qubit-qubit and photon-photon correlation function for different $g$ at $\lambda = -2$. (b) The concurrence of two qubits in a single cavity at various $g$ and $\lambda$. (c) The trace distance $\mathcal{T}$ defined by Eq (\ref{TraceDis1}) for various $g$ and $\lambda$. The red dashed line in (b) and (c) represents the boundary of Mott insulating and Superfluid phases}
\label{Fig3}
\end{figure}

\begin{figure*}[htp]
\centering
\includegraphics[scale=0.68]{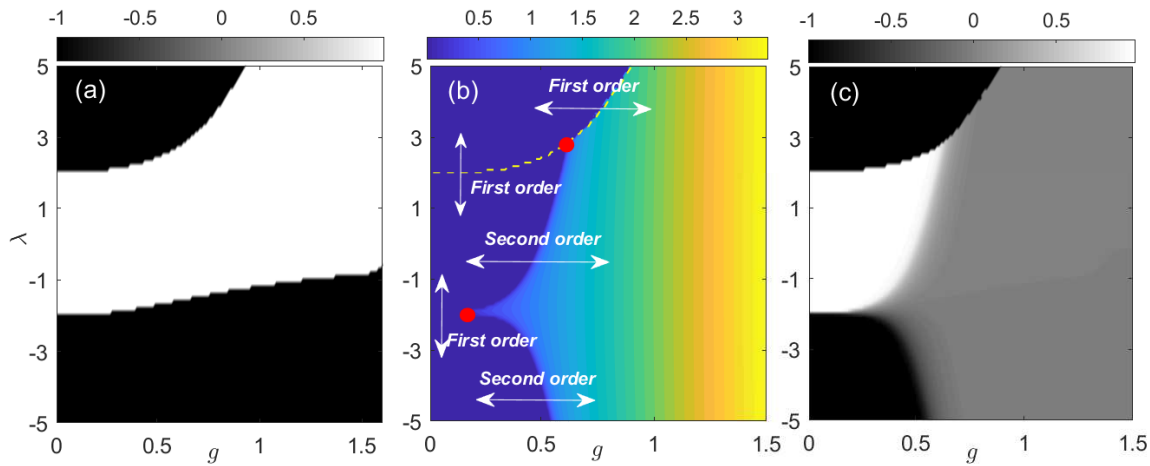}
\caption{
(a) The ground state mean value of the parity operator  $\langle\hat{\mathcal{P}}\rangle$ at $J = 0$ (b) The order parameter $ \psi$ for various $g$ and $\lambda$ at $J = 0.05$. The yellow dashed line separates regions with different angular momentum $\mathcal{J}$ as in Fig \ref{Fig1}.  (c) The ground state mean value of the parity operator  $\langle\hat{\mathcal{P}}\rangle$ at $J = 0.05$. $g_1=g_2=g$ for all figures above}
\label{Fig4}
\end{figure*}

In this subsection, we discuss some local and non-local properties of the ground state.  We first study the entanglement of two qubits in a single cavity by calculating the concurrence $\mathcal{C}_2 $ for various $\lambda$ and $g$ shown in Fig(\ref{Fig3}).b. The Mott-insulating phase can be divided into three regions. The ground state entanglement achieves its maximal in the region of  $\mathcal{N} = 1,\mathcal{J}=0$ and remains zero when $\mathcal{N}=0,\mathcal{J} =1$ due to the local reduced density matrix given in the last subsection. For the region with $\mathcal{N} = 1, \mathcal{J}=1 $, the local reduced density matrix  can be written as a mixture:
\begin{align}
 \rho_n =\sum_{i,j=1}^2 p_{ij}|\psi_{n,i}\rangle\langle \psi_{n,j}|
\end{align}
where $|\psi_{n,1}\rangle =|\downarrow\downarrow\rangle_{qb}|1\rangle_{ph} $ and $|\psi_{n,2}\rangle =(1/\sqrt{2})\left(|\uparrow\downarrow\rangle_{qb} + |\downarrow\uparrow\rangle_{qb}\right)|0\rangle_{ph}$.   The population of $p_{11}$ increases as $g$ increases, leading to the decrease of the concurrence.

As for the non-local property, we first calculate the spin-spin and photon-photon correlation function defined as $C_{\sigma_+\sigma_-}(R) = \langle \sigma_{+,i} \sigma_{-,i+R}\rangle$ and $ C_{a^\dag a}(R) =    \langle a^\dag_{i} a_{i+R}    \rangle$.  These correlation functions can be used to identify the Mott-insulating and Superfluid phases and have been extensively studied on the one-dimensional cavity lattice model in Ref\cite{flottat2016quantum,cui2019two,cui2020nonlinear}.  As shown in Fig \ref{Fig3}.(a), The power-law decay of $C_{\sigma_+\sigma_-}(R)$ and $ C_{a^\dag a}(R)$ at $g=1.2$ in the log-log plot  indicates the emergence of the  quasi-long range order and the superfluid phase. While for $g=0.5$, correlation functions display a faster decay in an exponential form, indicating the occurrence of the Mott-insulating phase.  Since the reduced density matrix for several adjacent cavities can be exactly obtained. We also investigate the correlation between two nearest cavities by calculating the trace distance defined as:
\begin{align}
&\mathcal{T} = \frac{1}{2}|| \rho_{M/2,M/2+1} - \rho_{M/2}\otimes \rho_{M/2} || \label{TraceDis1}
\end{align}
where $||\rho-\sigma || \equiv\text{Tr}\left[\sqrt{(\rho-\sigma)^\dag (\rho-\sigma) }\right] $ is the Schatten norm for $p=1$, $ \rho_{M/2,M/2+1}$  is the reduced density matrix of two adjacent sites in the center of the cluster and $\rho_{M/2}$ is the reduced density matrix of the single site in the middle.  The vanishing of $\mathcal{T}$  indicates that all cavities are decoupled and no correlation exist between cavities. As shown in  Fig(\ref{Fig3}).c,  $\mathcal{T}$ equals zero  in the region of  $\mathcal{N} = 1,\mathcal{J} = 0$ and $\mathcal{N}=0, \mathcal{J} = 1$. Thus the ground state there can be precisely  expressed within mean-field theory.  On the contrary, $\mathcal{T}$ achieves the maximum in the vicinity of the phase boundary due to the emergence of the critically. $\mathcal{T}$ in the superfluid phase is finite, suggests that qubits and photons from  two adjacent cavities are correlated, in accordance with  the power-law deay of various correlation functions as shown in Fig \ref{Fig3}. (a). Finally, the hopping effect cannot   produce the quantum entanglement between two qubits in different cavities as we calculate the concurrence but find no quantum entanglement in the whole parameter space.

\subsection{Effects of contour-rotating terms}

When $g_2\neq 0$, the U(1) symmetry of the Hamiltonian (\ref{H1}) is reduced to the $Z_2$ symmetry with the parity operator $\hat{\mathcal{P}} = \exp(i\pi \hat{\mathcal{N}})$.  To reveal the phase diagram in the current model, we plot $\psi$   as a function of $\lambda$ and $g$ by fixing $g_1 =g_2=g$ in Fig \ref{Fig4}.(b).  In analogy to Fig \ref{Fig1}, two types of quantum phase transitions are observed. Firstly, the quantum phase transition between the Mott-insulating and the Superfluid phase due to the $U(1)$ symmetry breaking is  replaced by the $Z_2$ symmetry-breaking phase transition between the incoherent phase and the coherent phase, as already observed in the Rabi-Hubbard model\cite{schiro2012phase,schiro2013quantum}. Note the non-integer value of the parity operator can also be used to characterize the occurrence of the coherent phase as shown in Fig \ref{Fig4}.(c). Similarly to the previous discussion for $g_2= 0$, this phase transition can be either the first-order or the second-order depending on whether the total angular momentum $\mathcal{J}$ varies across the phase boundary. Secondly, the incoherent phase is further divided into several regions due to its conserved parity $\mathcal{P} =\pm1$ with a first-order quantum phase transition between them. Finally, two triple points emerge in the phase diagram at $\lambda = 3$ and $\lambda =-2$ marked by red dots in Fig\ref{Fig4}.(b).

To further explore the local properties of different phases, we first calculate the concurrence $\mathcal{C}_2$ as a function of $\lambda$ and $g$ as shown In Fig \ref{Fig5}.(a). The existence of the qubit-qubit interaction assists in producing large entangled ground states. On the one hand, the maximal entanglement qubit state $|\psi_{qb}\rangle  = \frac{1}{\sqrt{2}}\left(|\uparrow\downarrow\rangle - |\downarrow\uparrow\rangle \right)$ is achieved due to the constrain of  $\mathcal{J}=0$ in the region of  $\mathcal{N} = 1,\mathcal{J} = 0$, similar as we found under the RWA.  On the one hand, the negative atom-atom interaction form a maximally entangled Bell state $|\psi_{qb}\rangle  = \frac{1}{\sqrt{2}}\left(|\uparrow\downarrow\rangle + |\downarrow\uparrow\rangle\right)$ at $g=0$ for $\lambda <-2$, leading to another region with large entanglement at $\lambda<-2$ when $g$ is small.

We also find that the non-local correlation in the coherent phase is  extremely suppressed by CRTs. To show that, we plot the trace distance $\mathcal{T}$ defined in Eq.(\ref{TraceDis1}) as a function of $g$ and $\lambda$ in Fig \ref{Fig5}.(b). Compared to Fig \ref{Fig3}.(c) where finite $\mathcal{T}$ is observed in the superfluid phase,  the trace distance rapidly decreases to zero in both the coherent and the incoherent phase away from the phase boundary, indicating that cavities are decoupled with each other. Thus  the ground state can be precisely described within the mean-field theory. As another example, we redefine the qubit and photon correlation function between two nearest cavities as $\tilde{C}_{\sigma_+,\sigma_-} = \langle \sigma_{+,i} \sigma_{-,i+1}\rangle - \langle \sigma_{+,i}\rangle \langle \sigma_{-,i+1}\rangle$  and $\tilde{C}_{a^\dag,a} =  \langle a^\dag_i  a_{i+1}\rangle- \langle a^\dag_i\rangle \langle a_{i+1}\rangle$.  Both correlation functions quickly decrease to zero as the increase of $g_2/g_1$.  This coincides with the result in Ref\cite{flottat2016quantum} where  correlation functions such as $C_{\sigma_+,\sigma_-}(R)$ and $C_{a^\dag,a}(R)$  reach plateaus with constant values even at large distance $R$ in the coherent phase in the Rabi-Hubbard model.

\begin{figure}[htp]
\centering
\includegraphics[scale=0.75]{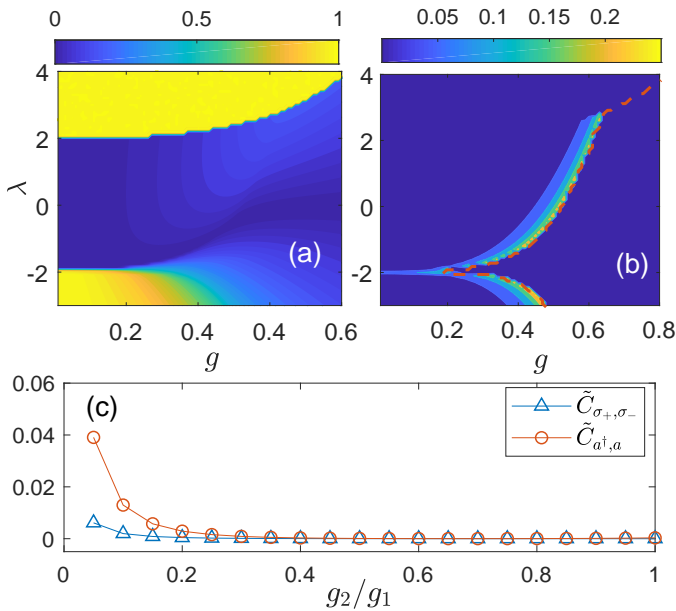}
\caption{ (a) The concurrence of two qubits in a single cavity for various $\lambda$ and $g$ with $g_1=g_2=g$.  (b) The trace distance $\mathcal{T}$  as a function of the $g$ and $\lambda$ with $g_1=g_2=g$. (c) The qubit and photon correlation functions vary with the  ratio $g_2/g_1$ at fixing $g_1 = 1.2$ and $\lambda = -1$. The red dashed in (b) represents the quantum phase boundary between the incoherent phase and the coherent phase.}
\label{Fig5}
\end{figure}

\section{Conclusion}

In this paper, we studied the ground state properties and the quantum phase transition of the Dicke-Hubbard model with XY qubit-qubit interaction. Using the exact numerical method combining the cluster mean-field theory and the matrix product, we illustrate the phase diagram and investigate the local and nonlocal properties of the ground state.

Under the RWA, we observed a transition from the Mott-insulating phase to the superfluid phase which can be either the first-order or the-second order depending on whether the total angular momentum changes across the phase boundary. Together with first-order phase transitions between different total excitation numbers, a rich phase diagram consisting of two quantum triple points emerge at both positive and negative $\lambda$.

As for the properties of the different phases, we first calculate the concurrence of two qubits in a single cavity and find that a large magnitude of qubit-qubit interaction assist in forming entanglement states. The maximum entanglement is observed in the region of  $\mathcal{N} = 1,\mathcal{J}=0$ and  $\mathcal{N} = 1,\mathcal{J}=1$. Then, we measure the non-local correlation between two nearest cavities through the trace distance $\mathcal{T}$ between the reduced density matrix of two adjacent sites and the product of the reduced density matrix of a single cavity. $\mathcal{T}$ reaches its maximum in the vicinity of the Mott-superfluid phase boundary. Finite correlation is observed in the superfluid phase, in accordance with the power-law decay of various correlation functions. Finally, we calculate the concurrence between a pair of qubits from two different nearest cavities, find no entanglement between them.

When CRTs are considered, the $U(1)$ symmetry is reduced to the $Z_2$  symmetry. The ground state undergoes a quantum phase transition between the incoherent phase and the coherent phase identified by the order parameter $\psi$ and the mean value of the parity $\mathcal{P}$. In analogous to the case $g_2=0$, the phase diagram consists of two quantum triple points associated with both first- and second-order coherent-incoherent phase transitions.  We also conclude that the non-local correlation in the coherent phase is extremely suppressed by CRTs as the trace distance $\mathcal{T}$ rapidly decreases with the increase of $g_2/g_1$ and remains zero in the coherent phase away from the phase boundary.

\section{Acknowledgement}
Shu He is supported by the National Natural Science Foundation of China under Grant No. 11804240. 11704093. Yan-Zhi Wang is supported by the National Natural Science Foundation of China under Grant No. 12105001 and Natural Science Foundation of Anhui Province under Grant No. 2108085QA24.  Chen Wang is supported by the National Natural Science Foundation of China under Grant Nos. 11704093 and 11547124.  Qing-Hu Chen is supported by the National Science Foundation of China under No. 11834005, the National Key Research and Development Program of China under No. 2017YFA0303002.

\section{Appendix:The numerical method}

In this Appendix,  we describe some details on the numerical method used in this study. In the cluster-mean-field theory, the correlation between lattice sites is assumed to be finite, namely the cluster size $M$. Thus the ground state of Hamiltonian (\ref{H1}) in the thermodynamic limit can be written as a product state:
\begin{align}
  |\psi_{GS}\rangle_{\text{CMF}} = \prod_c \otimes |\Psi_c(M)\rangle
\end{align}
where $|\Psi_c(M)\rangle$ is the ground state of the following cluster Hamiltonian:
\begin{align}
  H_{\text{cluster}} &= \sum_{m=1}^M H_{D,m} -J\sum_{m=1}^{M-1}(a_m^\dag a_{m+1} + a_m a^\dag_{m+1})\notag \\
  &-J(\psi_R a_1^\dag + \psi_R^*a_1 ) -J(\psi_L a_M^\dag + \psi_L^*a_M ) \label{Hc}
\end{align}

\begin{figure}[htp]
\centering
\includegraphics[scale=0.5]{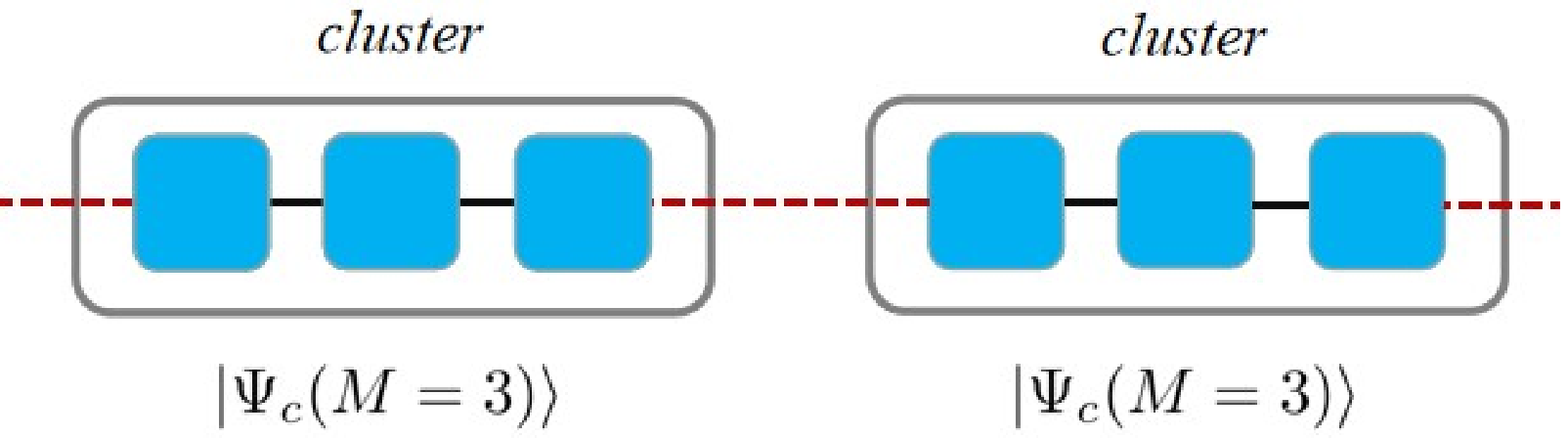}
\caption{The cluster mean field theory with cluster size $M =3$. The wave function of Hamiltonian (\ref{H1}) in the thermodynamic limit is approximated as product of ground states of each cluster $|\Psi_c\rangle$. The solid line between two adjacent lattice site represents the photon hopping interaction while the dashed line between clusters is the effective mean field interaction.}
\end{figure}

Note, the last two terms in (\ref{Hc}) are effective mean field interactions related to inter-cavity photon hopping. $\psi_L=\langle \Psi_c(M)|  a_1  |\Psi_c(M)\rangle$  is the expectation value of the boson annihilation operator for the leftmost site of the cluster while $\psi_R$ corresponding to the rightmost site. Obviously, the mean filed theory is recovered when $M=1$.  Then, the standard MPS method, which is highly efficient for one dimensional lattice model with nearest interaction \cite{schollwock2011density}, is applied to calculate the ground state of $(\ref{Hc})$. The matrix product operator form of the Hamiltonian (\ref{Hc}) can be written as:
\begin{align}
 & H_{\text{cluster}} = \prod_m^M H(m) \notag\\
 & H(1) =\left(
          \begin{array}{cccc}
            H_{D,1}-J(\psi_R a_1^\dag + \psi_R^*a_1 ) & a^\dag_m & a_m & I \notag\\
          \end{array}
        \right)\\
 & H(m\neq 1,M) =\left(
          \begin{array}{cccc}
            I &   &   &   \\
            -Ja_m &   &   &   \\
            -Ja^\dag_m &   &   &   \\
            H_{D,m} & a^\dag_m & a_m & I \notag\\
          \end{array}
        \right)\\
 &H(M) =\left(
          \begin{array}{cccc}
            H_{D,M}-J(\psi_L a_M^\dag + \psi_L^*a_M ) & a^\dag_m & a_m & I \notag\\
          \end{array}
        \right)^T\\
\end{align}
where $I$ is the identity operator. Finally, $\psi_L,\psi_R$ are updated iteratively until the self-consistency is achieved with appropriate chosen controlled-parameter such as the bond dimension of MPS $D$  the truncation of the photon number in each cavity $N_\text{tr}$. We also set the number of lattice sites $M$ to be large enough to eliminate the finite size effect.

As the first example to show the convergence performance of our numerical method, we calculate the order parameter $\psi$ defined in Eq.(\ref{psiCMF}) in the main text and the photon-photon correlation function between two adjacent cavities in the center of the cluster:
\begin{align}
\tilde{C}_{a^\dag a} = \langle a^\dag_{i} a_{i+1} \rangle - \langle a^\dag_{i}\rangle \langle a_{i+1} \rangle
\end{align}
where $\langle \sigma \rangle = \langle \Psi_c(M)|\sigma|\Psi_c(M)\rangle$ is the ground state mean value of the operator $\sigma$.
As shown in Fig(\ref{Fig6}), both  $\psi $ and  $C_{a^\dag a}$  reach convergence as we increase the tunable parameters. The ground states far away from the phase boundary converge quickly even we use $M = 6,D=4,N_{tr}=10$. While tens of cluster sites($M =64$) and larger truncations($D=12 ,N_{tr} =24$) are required to faithfully capture the critical behaviour close to the phase transition boundary.

\begin{figure}[htp]
\centering
\includegraphics[scale=0.75]{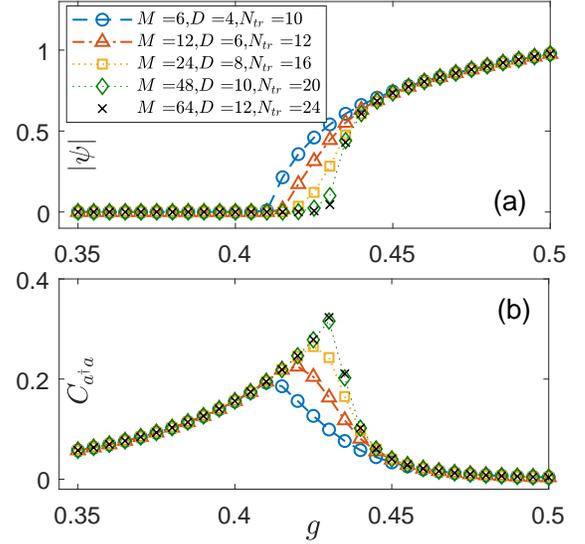}
\caption{ The Convergence of the numerical method. (a) The order parameter $ \psi$ as a function of $g$. (b) photon-photon correlation function between two adjacent cavities in the center of the cluster.  Other parameters are $\Delta =\omega = 1, J = 0.05, \lambda = -1, g_1=g_2=g$}
\label{Fig6}
\end{figure}

As the second example to demonstrates the excellent  performance of this method, we calculate the  the long-range correlation function defined as:
\begin{align}
 & C_{a^\dag a}(R) = \langle a^\dag_{M/2} a_{M/2+R} \rangle \\
 & C_{\sigma_+ \sigma_-}(R) = \langle \sigma_{+,{M/2}} \sigma_{-,{M/2+R}} \rangle
\end{align}

\begin{figure}[htp]
\centering
\includegraphics[scale=0.65]{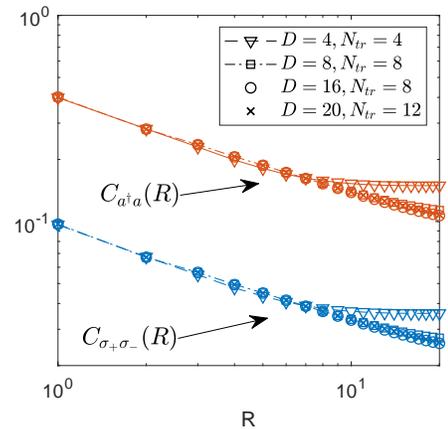}
\caption{ Dependence of correlation functions on the MPS bond dimension $D$ and the photon truncation number $N_{tr}$.  Other parameter are $\Delta = \omega = 1,  J = 0.01, \lambda = -2, g_1 =1.0, g_2=0, M = 72$}\label{Fig7}
\end{figure} Both $C_{a^\dag a}(R)$ and  $ C_{\sigma_+ \sigma_-}(R) $ should decay in power-law according as mentioned  in the main text. As shown in Fig(\ref{Fig7}), with sufficiently larege truncations($D=20, N_{tr}=20$), both $C_{a^\dag a}(R)$ and $C_{\sigma_+\sigma_-}(R)$  converge to a power-law decay and successfully capture the long-range correlations  up to $R \approx 20 $.  All results in the main text are carefully checked to reach convergence.

\newpage
\bibliographystyle{unsrt}%
\bibliography{XYLattice}

\end{document}